\documentclass[preprint]{elsarticle}
\biboptions{sort&compress}

\usepackage{graphicx}
\usepackage{amsmath,amssymb}
\usepackage{hyperref}
\usepackage{color}

\begin{document}
	
\title{Dirac fields in flat FLRW cosmology: Uniqueness of the Fock quantization}
	
\author[uam]{Jer\'onimo Cortez}
\ead{jacq@ciencias.unam.mx}
\author[iem]{Beatriz Elizaga Navascu\'es\corref{cor1}}
\ead{beatriz.elizaga@iem.cfmac.csic.es}
\author[ru]{Mercedes Mart\'in-Benito}
\ead{m.martin@hef.ru.nl}
\author[iem]{Guillermo A. Mena Marug\'an} \ead{mena@iem.cfmac.csic.es}
\author[cov]{Jos\'e M. Velhinho}
\ead{jvelhi@ubi.pt}

\cortext[cor1] {Corresponding author}

\address[uam]{Departamento de F\'isica, Facultad de Ciencias, Universidad Nacional Aut\'onoma de M\'exico, M\'exico D.F. 04510, M\'exico}
\address[iem]{Instituto de Estructura de la Materia, IEM-CSIC, Serrano 121, 28006 Madrid, Spain}
\address[ru]{Radboud University Nijmegen, Institute for Mathematics, Astrophysics and Particle Physics, Heyendaalseweg 135, NL-6525 AJ Nijmegen, The Netherlands}
\address[cov]{Universidade da Beira Interior, Rua Marqu\^es d'\'Avila e Bolama, 6201-001, Covilh\~a, Portugal}

\begin{abstract}

We address the issue of the infinite ambiguity that affects the construction of a Fock quantization of a Dirac field propagating in a cosmological spacetime with flat compact sections. In particular, we discuss a physical criterion that restricts to a unique possibility (up to unitary equivalence) the infinite set of available vacua. We prove that this desired uniqueness is guaranteed, for any possible choice of spin structure on the spatial sections, if we impose two conditions. The first one is that the symmetries of the classical system must be implemented quantum mechanically, so that the vacuum is invariant under the symmetry transformations. The second and more important condition is that the constructed theory must have a quantum dynamics that is implementable as a {(non-trivial)} unitary operator in Fock space. Actually, this unitarity of the quantum dynamics leads us to identify as explicitly time dependent some very specific contributions of the Dirac field. In doing that, we essentially characterize the part of the dynamics governed by the Dirac equation that is unitarily implementable. The uniqueness of the Fock vacuum is attained then once a physically motivated convention for the concepts of particles and antiparticles is fixed.

\end{abstract}

\begin{keyword}
Quantum Field Theory in Curved Spacetimes \sep Fock quantization \sep Uniqueness criteria \sep Unitarity in cosmological backgrounds

\PACS 04.62.+v \sep 03.70.+k \sep 98.80.Qc \sep 04.60.-m
\end{keyword}

\maketitle

\section{Introduction}
\label{sec:Intro}

Over the last half century, cosmology has developed into a solid scientific discipline experimenting an outstanding progress, both from the observational and theoretical points of view. Indeed, since the Cosmic Microwave Background (CMB) was discovered in 1965 \cite{cmb1,cmb2}, the series of accumulated technological advances has made possible the acquisition of very precise data, confirming the flatness of the expanding Universe and providing details about the anisotropies that are present in the temperature of this primordial radiation \cite{wmap,planck}. Such anisotropies can be traced back to small deviations from homogeneity and isotropy in the Early Universe, which also acted as seeds for the formation of the structures that we see nowadays at large scales \cite{structures}. These observations on the CMB have been theoretically contextualized, to a high level of accuracy, in the framework of cosmological perturbations described in General Relativity. Within this framework, the available data support the idea that these perturbations were originated during an inflationary period in the first epochs of our Universe \cite{pert1,pert2}. The standard approach is to describe those perturbations quantum mechanically, using the formalism of quantum field theory. In particular, this implies adopting a Fock representation for the (field analogue of the) Weyl algebra of the perturbation fields \cite{wald}. To arrive at predictions for the CMB, the relevant inhomogeneities are typically those corresponding to bosonic types of matter, as it is the case of scalar and tensor perturbations \cite{pert1,pert2}. Despite the proven success of this approach, most of the known fundamental theories involve spinorial degrees of freedom, as Dirac fermions are expected to be basic constituents of (at least part of) the matter content in our Universe. Therefore, it is interesting to incorporate this type of fields in the inflationary paradigm, for instance in order to see whether their evolution during those stages of the early history of the Universe may have left any trace, with potential consequences for the latter evolution of matter.

Actually, the possible relevance of free fermion fields as (test) perturbations propagating in a flat Friedmann-Lema\^itre-Robertson-Walker (FLRW) cosmology, both during the inflationary period and outside of it, has been addressed in several works over the last decades (see e.g. \cite{cosmof1,cosmof2,cosmof3,cosmof4,cosmof5,cosmof6,cosmof7,cosmof8}). An important issue in those investigations is the fundamental criteria employed for the choice of a Fock representation of the canonical anticommutation relations (CARs) of the Dirac field. In fact, the used criteria are somewhat vague, and the selected vacua are, if at all, typically motivated from adiabatic conditions \cite{parker}. Given this lack of universally accepted criteria, applicable in any flat FLRW spacetime, the problem stands that there may exist choices of Fock representations that, while still motivated by arguably valid physical arguments, lead to quantum theories that are not unitarily equivalent \cite{wald}. This ambiguity, inherent to quantum field theories in curved backgrounds, is specially severe in non-stationary spacetimes such as the cosmology addressed here. In fact, time-translational symmetry happens to be enough to select a unique Fock quantum vacuum if one imposes it quantum mechanically and requires the positiveness of the (associated concept of) energy \cite{kayb,baez}; however such symmetry is absent in non-stationary backgrounds. With this motivation, the aim of this work is demonstrating that, in flat FLRW cosmology with compact spatial sections, there actually exists a well defined criterion --composed of two conditions-- that selects a unique equivalence class of Fock representations of the CARs of a free minimally-coupled Dirac field. These conditons are the following. a) The Fock vacuum must be invariant under the group of (spatial and spinorial) symmetries of the field in the studied cosmology. b) The quantum dynamics of the Dirac field must be implementable as a unitary operator on Fock space \cite{unit} (once a precise explicit time dependence of the field is extracted). This latter requirement is actually the least that one would expect in order to avoid loss of information at the level of the quantum evolution of the system.

The generality of the criterion formed by these conditions can already be justified by its success to ensure the uniqueness of the Fock representation of a free real Klein-Gordon field in a variety of geometries \cite{gowdy1,gowdy2,gowdy3,gowdy4,gowdy5,gowdy6,gowdy7,compact1,compact2,compact3}, including the flat cosmology with compact spatial sections considered here \cite{torus1,torus2}. In fact, the requirement of a unitarily implementable dynamics permits to single out a unique canonical pair for the field (i.e, the field variable and its momentum) among all those related through global time-dependent rescalings \cite{uniqmom}. In other words, the requirement serves to characterize the part of the Klein-Gordon evolution that can be unitarily implemented, among those dynamics related by a field rescaling with a global function of time. Moreover, the condition of implementable dynamics in Fock space, combined with vacuum invariance under the group of background isometries, has proven to select a unique Fock representation (up to unitary equivalence) for the Dirac field propagating in an FLRW cosmology with spherical spatial sections \cite{uf1,uf2}, a model that was first studied in \cite{H-D}. Once again, the requirement that the theory admit a unitarily implementable dynamics fixed (in an essentially unique way) the part of the Dirac evolution that preserves quantum coherence. This was done by extracting from the field certain time-dependent contributions, regarded as background functions, and which display as well a dependence on the mass. Finally, the uniqueness criterion has been also proven to succeed for the Fock quantization of Dirac fields propagating in conformally ultrastatic backgrounds in 2+1 dimensions \cite{uf3}.

The aim of this work is to extend the uniqueness result of \cite{uf2} to the more realistic scenario of a free Dirac field propagating in an FLRW cosmology with compact spatial sections that are flat (instead of positively curved), namely three-dimensional tori ($T^{3}$). In order to attain such uniqueness, the symmetry that we impose on the Fock vacuum ensures invariance under the group of Killing isometries of $T^{3}$. In addition, we take into account the fact that the helicity of the Dirac field is a conserved quantity, and hence we impose as well the invariance of the vacuum under the rotation group generated by it. On the other hand, the condition of a unitarily implementable dynamics turns out to specify again what part of the fermion field evolves quantum mechanically (in a Heisenberg picture), both in the case of a massive and of a massless field. It is worth mentioning that the uniqueness (up to unitary equivalence) of the Fock representation is obtained once a convention for the notions of particles and antiparticles is set, as it also happened in \cite{uf2}.

The paper is structured as follows. In Sec. \ref{sec:model} we introduce the FLRW-Dirac system. The purpose of Sec. \ref{sec:quantization} is to characterize those Fock representations of the CARs of the Dirac field that satisfy the two conditions of our criterion. We first determine the vacua that are invariant under the group of Killing isometries of the spatial sections, and under the group generated by the helicity. Then, we select those annihilation and creation-like variables that lead to representations which allow for a unitarily implementable (and non-trivial) dynamics. In Sec. \ref{sec:uniqueness} we prove the uniqueness of the family of vacua selected by our criterion, once a convention for the notions of particles and antiparticles has been given. Besides, the requirements for a physically reasonable convention are discussed. Finally, in Sec. \ref{sec:conclu} we summarize our results and conclude.   

\section{The Dirac field in flat cosmology}
\label{sec:model}

The system that we will consider in this work is a Dirac field minimally coupled to a homogeneous and isotropic spacetime with flat spatial sections that are isomorphic to the three-torus $T^{3}$. The metric can be written as 
\begin{equation}
ds^{2}=e^{2\alpha(\eta)}(-\text{d}\eta^{2}+h_{ij}\text{d}\theta^{i}\text{d}\theta^{j}),
\end{equation} 
with $i,j=1,2,3$, and where we have set the lapse as corresponding to conformal time, $\eta$. Besides, $h_{ij}$ denotes the standard flat metric on $T^{3}$ and we have chosen spatial orthogonal coordinates $\theta_{i}$ such that $2\pi\theta_{i}/l_{0}\in S^{1}$, where $l_{0}$ is the compactification period in each of the orthogonal directions. This cosmological space-time clearly admits a global system of orthonormal tetrads. Therefore, its bundle of orthonormal and oriented frames is trivial, and a spin structure can be defined on it \cite{Geroch,SGeom}. In order to keep the discussion as general as possible, we will not fix any preferred spin structure. The Dirac fields $\Psi$ of mass $m$ may be regarded as the cross-sections of the resulting spinor bundle, such that they obey the following dynamical equation with well-posed Cauchy problem \cite{Dimock}:
\begin{align}\label{Deq}
e^{\mu}_{a}\gamma^{a}\nabla^{S}_{\mu}\Psi+m\Psi=0.
\end{align}
Here $\mu=0,1,2,3$ is a world index, whereas $a=0,1,2,3$ is an internal gauge index. Besides,  $\nabla^{S}_{\mu}$ refers to the spin lifting of the Levi-Civit\`{a} covariant derivative \cite{SGeom}, $e^{\mu}_{a}$ denotes the tetrad, and $\gamma^{a}$ are the Dirac matrices, that generate the Clifford algebra of a flat Lorentzian manifold of dimension four. For convenience in the following analysis, we take these generators in the so-called Weyl representation, as done in \cite{uf2}. Then, we can describe any Dirac spinor $\Psi$ by means of a pair of two-component spinors $\phi^{A}$ and $\bar{\chi}_{A'}$ (with $A=1,2$ and $A'=1',2'$), possessing well-defined and opposite chirality. In particular, $\phi^{A}$ is taken to be the left-handed projection of $\Psi$, whereas $\bar{\chi}_{A'}$ is the right-handed one. Throughout this paper, overbars will denote complex conjugation. In order to account for the anticommuting nature of the fermion fields, we will take the components of these spinors to be Grassmann variables \cite{Berezin}. Using the same spinor conventions as in \cite{H-D}, spinor indices are raised and lowered with the antisymmetric symbols $\epsilon^{AB}$, $\epsilon_{AB}$, $\epsilon^{A'B'}$, and $\epsilon_{A'B'}$, all of them represented by the matrix 
$$\begin{pmatrix}
0 & 1 \\ -1 & 0
\end{pmatrix}.$$
Thus, we have $\phi_{A}=\phi^{B}\epsilon_{BA}$, ${\bar{\chi}}^{A'}=\epsilon^{A'B'}{\bar{\chi}}_{B'}$, etc.

Both from a Hamiltonian point of view \cite{T-N}, as well as to replace the spatial dependence of the Dirac equation \eqref{Deq} with a spectral collection of time-dependent coefficients that decouple dynamically, it is most convenient to perform now a partial gauge fixing. Specifically, the gauge group of the bundle of orthonormal and oriented frames can be reduced from $SO(3,1)$ (orthocronus) to $SO(3)$ \cite{Isham}. In the considered space-time, this procedure gives rise to a well-defined restriction of the spin structure to the double covering of the reduced bundle, which in turn can be understood so as to provide a spin structure on each of the three-tori that foliate the $4$-dimensional manifold. The two-component spinor fields introduced above are then one-parameter families of cross-sections of the resulting spinor bundle on $T^{3}$, with  parameter given by the conformal time $\eta$. In practice, we achieve this reduction by imposing the gauge-fixing condition known as ``time-gauge'', namely $e^{j}_{0}=0$. Once this is done, the only non-vanishing Dirac brackets \cite{Dirac} at time $\eta$ are given by (see \cite{T-N})
\begin{align} \{\Psi^{\dagger}(\eta,\vec{\theta}),\Psi(\eta,\vec{\theta}')\}=-ie^{-3\alpha(\eta)}\delta(\vec{\theta}-\vec{\theta}')I,
\end{align}
where the dagger denotes Hermitian conjugate, $I$ is the $4\times 4$ identity matrix, $\vec{\theta}:=(\theta_{1},\theta_{2},\theta_{3})$, and $\delta(\vec{\theta})$ is the Dirac delta on $T^{3}$. In addition, owing to the Grassmann nature of the spinors, these brackets are symmetric \cite{Casal}. The symplectic structure from which they arise then naturally defines a conserved inner product on the linear space of solutions $\mathcal{S}$ of the Dirac equation,
\begin{align}\label{inner}
(\Psi_{1},\Psi_{2})_{\mathcal{S}}=e^{3\alpha}\int \text{d}^{3}\vec{\theta}\,\Psi_{1}^{\dagger}\Psi_{2}, \qquad \Psi_{1},\Psi_{2}\in\mathcal{S}.
\end{align}

We will now use the spectral properties of the Dirac operator on $T^{3}$ to decompose the Dirac field in modes, and express the field equation as a set of equations of motion for the time-dependent coefficients of the mode expansion. Since the Riemannian manifold $T^{3}$ is geodesically complete, the Dirac operator on it is essentially self-adjoint in the inner product introduced above \cite{SGeom}. Its spectrum is discrete and well-known for any of the possible spin structures on $T^{3}$. It is real with eigenvalues given by (see \cite{Dtorus})
\begin{align}\label{eigenv}
\pm\omega_{k}=\pm \frac{2\pi}{l_{0}} \left|\vec{k}+\frac{1}{2}\sum_{j=1}^{3}\epsilon^{j}\vec{v}_{j}\right|.
\end{align}
Here, $\vec{k}\in\mathbb{Z}^{3}$ is any tuple of integers, the tuples $\vec{v}_{j}$ denote the standard orthonormal basis of the lattice $\mathbb{Z}^{3}$, and the three numbers $\epsilon^{j}\in\{0,1\}$ characterize each of the possible choices of spin structure on $T^{3}$. Given one of these spin structures, we identify the label $k$ in $\omega_{k}$  (or, equivalently, in $-\omega_{k}$) with the norm of any of the tuples $\vec{k}$ for which we get this Dirac eigenvalue. The number of tuples that correspond to this same value of $\omega_{k}$ (or $-\omega_k$) is the degeneracy $g_{k}$, for which no closed expression is known. Nonetheless, it is a well-known result of Riemannian geometry that $g_{k}$ is asymptotically larger than $\omega_k$ but negligible compared to $\omega_{k}^{3}$ in the regime of large positive eigenvalues $\omega_{k}$ \cite{JRoe}. 

Taking any valid spin structure on $T^{3}$, and any specification of triads, the eigenspinors of the Dirac operator then form a complete orthogonal basis [in the inner product \eqref{inner}] for the expansion of any two-component spinor on the spatial sections of the studied cosmology. In particular, if one chooses triads such that the spin connection vanishes, then the Dirac operator reduces to the usual one for a flat Euclidean manifold. Hence, its mutually orthogonal eigenspinors of the chirality, e.g., of $\phi^{A}$, with eigenvalues $\pm\omega_{k}$, have the form (omitting spinor indices)
\begin{align}\label{eigens}
u^{(\pm)}_{\vec{k}}\exp\left[{i\frac{2\pi}{l_{0}}\bigg( \vec{k}+\frac{1}{2}\sum_{j=1}^{3}\epsilon^{j}\vec{v}_{j}\bigg)\cdot\vec{\theta}}\right], 
\end{align}
where $u^{(\pm)}_{\vec{k}}$ are some $\vec{\theta}$-independent two-component spinors. They are subject to the condition that the eigenvalue equation must hold. Besides, they can be normalized so that $u^{(\pm)\dagger}_{\vec{k}}u^{(\pm)}_{\vec{k}}=1$, and they satisfy
\begin{align}\label{norm}
u^{(\pm)}_{\vec{k}'A}\epsilon^{AB}u^{(\pm)}_{\vec{k}B}=e^{iC^{(\pm)}_{\vec{k}}}\delta_{\vec{k}',-\vec{k}-\epsilon^{j}\vec{v}_{j}}, \qquad u^{(+)}_{\vec{k}'A}\epsilon^{AB}u^{(-)}_{\vec{k}B}=0,
\end{align}
for all $\vec{k},\vec{k}'\neq -\epsilon^{j}\vec{v}_{j}/2$. Summation over repeated indices is assumed here and from now on\footnote{Except for the index $\vec{k}$ on the right hand side of the first relation in \eqref{norm}}. Finally, the constants $C^{(\pm)}_{\vec{k}}$ are some phases that can be chosen conveniently by modifying those of $u^{(\pm)}_{\vec{k}}$. 

Since the eigenspinors \eqref{eigens} form a complete basis, we can express any spinor of the chirality of $\phi^{A}$ on $T^{3}$ in terms of them, independently of the choice of gauge. On the other hand, the complex conjugate of \eqref{eigens} provides a complete basis for the expansion of any spinor of the chirality of $\bar{\chi}_{A'}$. Let us call $m_{\vec{k}}$ and $\bar{r}_{\vec{k}}$ the time-dependent coefficients that multiply respectively $u^{(+)}_{\vec{k}}$ and $u^{(-)}_{\vec{k}}$ in the expansion of $e^{3\alpha/2}\phi^{A}$. Analogously, we will denote by $\bar{s}_{\vec{k}}$ and $t_{\vec{k}}$ the time-dependent coefficients that multiply respectively $\bar{u}^{(+)}_{\vec{k}}$ and $\bar{u}^{(-)}_{\vec{k}}$ in the expansion of $e^{3\alpha/2}\bar{\chi}_{A'}$. Then, recalling relations \eqref{norm}, the dynamical Dirac equation \eqref{Deq} for any choice of spin structure can be rewritten as the following set of equations for the introduced coefficients
\begin{align}\label{1order}
x_{\vec{k}}'=i\omega_{k}x_{\vec{k}}-ime^{\alpha}\bar{y}_{-\vec{k}-\epsilon^{j}\vec{v}_{j}}, \qquad \bar{y}_{\vec{k}}'=-i\omega_{k}\bar{y}_{\vec{k}}-ime^{\alpha}x_{-\vec{k}-\epsilon^{j}\vec{v}_{j}}, 
\end{align}
for all $\vec{k}\neq -\epsilon^{j}\vec{v}_{j}/2$. Here the prime stands for the derivative with respect to conformal time, and $(x_{\vec{k}},\bar{y}_{\vec{k}})$ denotes any of the ordered pairs $(m_{\vec{k}},\bar{s}_{\vec{k}})$ or $(t_{\vec{k}},\bar{r}_{\vec{k}})$, since they obey the same dynamics. We have excluded from the analysis the zero-modes of the decomposition of $\Psi$ (those corresponding to $\omega_{k}=0$), that only exist with the trivial spin structure on $T^3$. These modes can be isolated from the rest and their behavior will be irrelevant for our future considerations about the unitary implementability of Bogoliubov transformations. Finally, let us 
note that the above mode equations do not depend on the specific value of the tuple $\vec{k}$, but only on the corresponding Dirac eigenvalue (in norm). Therefore, as differential equations, they have the same structure as those analyzed in \cite{uf1}. We will refer to the results of that reference whenever needed.

\section{Criterion for the choice of vacua}
\label{sec:quantization}

In this section we will characterize those Fock representations of the CARs of the Dirac field that satisfy the two conditions that form our fundamental criterion. First of all, they must be such that the resulting vacua be invariant under the Killing symmetries of the toroidal spatial sections, as well as under the action of the rotation group generated by the helicity. Secondly, and more importantly, the Fock representation must admit dynamics implementable as unitary (and non-trivial) quantum transformations on Fock space.

In order to achieve this characterization, it will be most useful to introduce a mathematical object that fully determines each of the infinitely many Fock representations of the Dirac field: the complex structure \cite{wald}. In the covariant picture, this is a real linear map $\mathcal{J}:\mathcal{S}\rightarrow\mathcal{S}$ with a square that equals minus the identity, and such that it must be compatible with the symplectic structure of the system, or equivalently with the inner product \eqref{inner}: $(\mathcal{J}\Psi_{1},\mathcal{J}\Psi_{2})_{\mathcal{S}}=(\Psi_{1},\Psi_{2})_{\mathcal{S}}$. Once a complex structure has been chosen, the space of solutions $\mathcal{S}$ splits into its two eigenspaces of $\pm i$ eigenvalue. The completion in the product \eqref{inner} of the $+i$ eigenspace is taken to be the one-particle Hilbert space. In a completely analogous way, the complex structure $\mathcal{J}$ may be defined on the complex conjugate space $\bar{\mathcal{S}}$, with inner product equal to the complex conjugate of \eqref{inner}. The completion of the corresponding $+i$ eigenspace provides the one-antiparticle Hilbert space. The antisymmetric Fock space for the representation of the CARs is constructed then out of the direct sum of the one-particle and one-antiparticle Hilbert spaces. From this construction, any Dirac field $\Psi$ is represented by an infinite collection of particle annihilation and antiparticle creation operators, in a way that clearly depends on the choice of complex structure, that is, on the choice of the one-particle and one-antiparticle sectors. Furthermore, owing to the infinite amount of degrees of freedom present in the field, the different possible representations need not be unitarily equivalent to each other. We will remove this ambiguity by imposing the conditions mentioned above.

\subsection{Quantum implementation of the symmetries}

The vacuum of a Fock representation of the CARs of the Dirac field is defined as the cyclic state of the Fock space which vanishes under the action of each of the annihilation operators (both for particles and antiparticles). We want to restrict our attention to representations such that the Killing symmetries of the flat sections of our cosmological model, as well as the rotations generated by the helicity, can all be implemented as unitary operators that leave the vacuum state invariant. In other words, we want to restrict our discussion to complex structures that commute with 
the action of the group formed by these physical symmetries.

In order to specify the complex structures of interest, let us start by studying Killing symmetries. Since the spatial sections are (isomorphic to) $T^{3}$, we are going to consider the symmetry group formed by the composition of rigid rotations in each of the three periodic orthogonal directions, namely the composition of $T_{\alpha_{i}}:\theta_{i}\rightarrow\theta_{i}+\alpha_{i}$ with $2\pi\alpha_{i}/l_{0}\in S^{1}$. We will use the compact notation $T_{\vec{\alpha}}=T_{\alpha_{1}}\circ T_{\alpha_{2}}\circ T_{\alpha_{3}}$ for these compositions. For each choice of spin structure, it is clear that active $T_{\vec{\alpha}}$ transformations correspond to multiplication by a factor $e^{i2\pi\vec{k}\cdot\vec{\alpha}/l_{0}}e^{i\pi\epsilon^{j}\vec{v}_{j}\cdot\vec{\alpha}/l_{0}}$ in each of the elements \eqref{eigens} of the basis of eigenspinors of the Dirac operator. The second phase in this factor is in fact the same for all Dirac modes, and only depends on the choice of spin structure. Therefore, our group of Killing symmetries acts as a direct sum of irreducible representations of the unitary Abelian group $U(1)\times U(1)\times U(1)$ on the space of two-component spinors of a given chirality. For each tuple $\vec{k}\in\mathbb{Z}^{3}$, we get in this way two copies of a one-dimensional complex irreducible representation. It is also clear that different tuples $\vec{k}$ give rise to inequivalent representations of the group. A parallel analysis can be carried out for the complex conjugate elements of \eqref{eigens}, which provide a complete basis for the two-component spinors of opposite chirality. Specifically, under the action of the considered symmetry group, each of the eigenspinors corresponding to a given $\vec{k}$ gets now multiplied by $e^{-i2\pi(\vec{k}+\epsilon^{j}\vec{v}_{j})\cdot\vec{\alpha}/l_{0}}e^{i\pi\epsilon^{j}\vec{v}_{j}\cdot\vec{\alpha}/l_{0}}$. Then, recalling our mode expansion of the Dirac field $\Psi$ and making use of Schur's lemmas \cite{Reps}, we conclude that any complex structure $\mathcal{J}$ on $\mathcal{S}$ that commutes with the unitary representation of our group of Killing symmetries can at most mix the coefficients $(m_{\vec{k}},\bar{s}_{-\vec{k}-\epsilon^{j}\vec{v}_{j}},t_{-\vec{k}-\epsilon^{j}\vec{v}_{j}},\bar{r}_{\vec{k}})$ among them, for each tuple $\vec{k}\in\mathbb{Z}^{3}$.

An additional symmetry of the Dirac-FLRW system, that we include in our discussion, arises from the conservation of the helicity of the fermion field in the evolution with respect to the conformal time. Indeed, in the context of a free fermion particle propagating in the studied cosmological spacetime, one may consider the projection of the spin angular momentum in the direction of the linear three-momentum of the particle, since the spatial sections are flat. This operator defines the helicity, namely:
\begin{align}
\mathfrak{h}=[-\vec{\nabla}^{2}]^{-1/2}\begin{pmatrix} -i\vec{\Sigma}\cdot\vec{\nabla} & 0 \\ 0 & -i\vec{\Sigma}\cdot\vec{\nabla} \end{pmatrix},
\end{align}
where $\vec{\nabla}$ is the standard gradient with respect to the three orthogonal coordinates $\theta_{i}$, and $\vec{\Sigma}$ is a compact notation for the three Pauli matrices. This operator is only well-defined in the closed subspace of $\mathcal{S}$ (and of its complex conjugate) formed by the linear span of all the pairs of eigenspinors \eqref{eigens} corresponding to tuples $\vec{k}\in\mathbb{Z}^{3}$ such that $\omega_{k}\neq 0$. We will say that a fermion field has positive or negative helicity when it is an
eigenspinor of $\mathfrak{h}$ with eigenvalue +1 or -1, respectively.

Let us note that, in the gauge of a vanishing spin connection, the matrix blocks of the helicity operator $\mathfrak{h}$ correspond precisely to the Dirac operator on $T^{3}$. Then, it is not difficult to see that the part of positive helicity of the Dirac field $\Psi$ is that spanned by the coefficients $m_{\vec{k}}$ and $\bar{s}_{\vec{k}}$ for all $\vec{k}\in\mathbb{Z}^{3}$ different from $-\epsilon^{j}\vec{v}_{j}/2$. On the other hand, the negative helicity contribution is the part spanned by all the coefficients $t_{\vec{k}}$ and $\bar{r}_{\vec{k}}$. By inspection of the equations of motion \eqref{1order}, one can easily check that the helicity is a quantity conserved under evolution in conformal time. We can then consider as an additional symmetry of the system the one-parameter group of spin rotations generated by the helicity via complex exponentiation of $\mathfrak{h}/2$ times the angle of the rotation. This group is immediately unitary in the inner product \eqref{inner}, given that $\mathfrak{h}$, constructed out of the Dirac operator, is essentially self-adjoint. To implement the symmetry naturally in the quantum theory, it suffices then to notice that any complex structure that commutes with it cannot mix the time-dependent coefficients that correspond to positive helicity with those that describe the part of negative helicity of the Dirac field.

Summarizing, any complex structure $\mathcal{J}$ that commutes with the action of the discussed Killing symmetries of the spatial sections, as well as with the action of spin rotations generated by the helicity of the Dirac field, must display a $2\times 2$ block structure. More concretely, the blocks in which $\mathcal{J}$ must decompose can at most mix the pairs of coefficients $(m_{\vec{k}},\bar{s}_{-\vec{k}-\epsilon^{j}\vec{v}_{j}})$ or $(t_{\vec{k}},\bar{r}_{-\vec{k}-\epsilon^{j}\vec{v}_{j}})$, for each possible $\vec{k}$. In what follows, we will call invariant any complex structure with these properties, as well as the Fock representation determined by it.

\subsection{Conditions for unitary dynamics}

It follows from our previous analysis that, at any time $\eta$, an invariant complex structure defines a family of annihilation and creation-like variables of the general form
\begin{align}\label{anni}
a_{\vec{k}}^{(x,y)}(\eta)=f_{1}^{\vec{k}}(\eta)x_{\vec{k}}(\eta)+f_{2}^{\vec{k}}(\eta)\bar{y}_{-\vec{k}-\epsilon^{j}\vec{v}_{j}}(\eta), \\ \label{creat}
b_{\vec{k}}^{\dagger (x,y)}(\eta)=g_{1}^{\vec{k}}(\eta)x_{\vec{k}}(\eta)+g_{2}^{\vec{k}}(\eta)\bar{y}_{-\vec{k}-\epsilon^{j}\vec{v}_{j}}(\eta),
\end{align}
together with their complex conjugates. Here, $a_{\vec{k}}^{(x,y)}$ and $b_{\vec{k}}^{(x,y)}$ denote, respectively, particle and antiparticle annihilation-like variables, whereas
\begin{equation} a_{\vec{k}}^{\dagger (x,y)}:=\bar{a}_{\vec{k}}^{(x,y)}, \quad b_{\vec{k}}^{\dagger (x,y)}:=\bar{b}_{\vec{k}}^{(x,y)} 
\end{equation}
are the corresponding creation-like variables. One can express the Dirac field $\Psi$ in terms of these variables. Recall that $(x_{\vec{k}},\bar{y}_{\vec{k}})$ denotes either $(m_{\vec{k}},\bar{s}_{\vec{k}})$ or $(t_{\vec{k}},\bar{r}_{\vec{k}})$. Besides, the complex time-dependent functions $f_{l}^{\vec{k}}$ and $g_{l}^{\vec{k}}$, with $l=1,2$, as well as their complex conjugates, are taken to be as smooth as needed for a well-posed evolution of the annihilation and creation-like variables. In addition, they satisfy
\begin{align}\label{sympl}
\left\vert f_{1}^{\vec{k}}\right\vert^{2}+\left\vert f_{2}^{\vec{k}} \right\vert^{2}=1, \qquad \left\vert g_{1}^{\vec{k}}\right\vert^{2}+\left\vert g_{2}^{\vec{k}}\right\vert^{2}=1, \qquad f_{1}^{\vec{k}}\bar{g}^{\vec{k}}_{1}+f_{2}^{\vec{k}}\bar{g}^{\vec{k}}_{2}=0,
\end{align}
so that the variables that they define indeed have canonical Dirac brackets that remain constant in time \cite{uf1,uf2}. These conditions guarantee that relations \eqref{anni} and \eqref{creat} are invertible, and allow us to write
\begin{align}\label{fgrel}
g_{1}^{\vec{k}}=\bar{f}_{2}^{\vec{k}}e^{iG^{\vec{k}}}, \qquad g_{2}^{\vec{k}}=-\bar{f}^{\vec{k}}_{1}e^{iG^{\vec{k}}},
\end{align}
with $G^{\vec{k}}$ some (possibly time-dependent) phase.

The dynamical evolution of the variables $x_{\vec{k}}$ and $\bar{y}_{\vec{k}}$ from an arbitrary initial time $\eta_{0}$ to any other time $\eta$, which is dictated by \eqref{1order},  may be regarded as a linear transformation of the corresponding initial data at $\eta_{0}$ \cite{uf1}. Using such transformation, one can obtain the relation between the values of the annihilation and creation-like variables at any time $\eta$ and their initial values at $\eta_{0}$. This latter relation is just a Bogoliubov transformation. Let us call $\beta_{\vec{k}}^{f}(\eta,\eta_0)$ and $\beta_{\vec{k}}^{g}(\eta,\eta_0)$ the coefficients of the antilinear part of this transformation, which respectively relate $a_{\vec{k}}^{(x,y)}(\eta)$ with $b_{\vec{k}}^{\dagger (x,y)}(\eta_{0})$, and $b_{\vec{k}}^{\dagger (x,y)}(\eta)$ with $a_{\vec{k}}^{(x,y)}(\eta_{0})$. In \cite{uf1} it was shown that their norm is given by
\begin{align}\label{beta}
|\beta_{\vec{k}}^{h}(\eta,\eta_0)|=&\Bigg|\left[-h_{1}^{\vec{k}}\bigg(h_{2}^{\vec{k},0}+\Gamma_k h_{1}^{\vec{k},0}\bigg)e^{i\int \Lambda^{1}_{k}}+\bar\Gamma_k h^{\vec{k}}_{2}h_{2}^{\vec{k},0}e^{\Delta\alpha} e^{i\int\bar{\Lambda}^{2}_{k}}\right] e^{i\omega_{k}\Delta\eta}\nonumber\\& +\left[h_{2}^{\vec{k}}\bigg(h_{1}^{\vec{k},0} -\bar\Gamma_k h_{2}^{\vec{k},0}\bigg)e^{-i\int \bar{\Lambda}^{1}_{k}}+\Gamma_k h^{\vec{k}}_{1}h_{1}^{\vec{k},0}e^{\Delta\alpha}e^{-i\int\Lambda^{2}_{k}}\right] e^{-i\omega_{k}\Delta\eta}\Bigg|,
\end{align}
where $h$ can be either $f$ or $g$, and we have introduced the notation $\Delta\eta=\eta-\eta_{0}$ and $\Delta\alpha=\alpha-\alpha_{0}$, with $\alpha_{0}=\alpha(\eta_{0})$. Also, we have omitted the dependence of the functions $h_l^{\vec{k}}$ on $\eta$, we have distinguished evaluation at $\eta_0$ with the superscript 0 (preceded by a comma), and we have defined
\begin{align}\label{gamma}
\Gamma_{k}=\frac{me^{\alpha_{0}}}{2\omega_{k}+i\alpha_{0}'},
\end{align}
where (as before) the prime stands for the derivative with respect to the conformal time. Besides, all integrals in \eqref{beta} are in conformal time, in the interval $[\eta_{0},\eta]$, and $\Lambda^{1}_{k}$ and $\Lambda^{2}_{k}$ denote two time-dependent functions of which we will only need to know their asymptotic behavior in the regime of large values of the Dirac eigenvalue $\omega_{k}$. The analysis of \cite{uf1} proves that, in fact, this behavior is of the order of the inverse of $\omega_{k}$, namely $\mathcal{O}(\omega_{k}^{-1})$, as long as the function $\alpha(\eta)$ and its derivatives (up to third order) exist and are integrable in every closed interval $[\eta_{0},\eta]$ covered in the evolution.

Given the arbitrary but fixed initial time $\eta_{0}$, let us call $\mathcal{J}_{\eta_{0}}$ the complex structure that selects as annihilation and creation-like variables those determined by \eqref{anni} and \eqref{creat} at the considered initial time $\eta_0$. Obviously, we can express the Dirac field $\Psi$ at any time in terms of these variables associated with the initial section at $\eta_{0}$. On the other hand, we can also express the pairs $a_{\vec{k}}^{(x,y)}(\eta_{0}),b_{\vec{k}}^{\dagger (x,y)}(\eta_{0})$ in terms of their time-evolved values $a_{\vec{k}}^{(x,y)}(\eta),b_{\vec{k}}^{\dagger (x,y)}(\eta)$ at any time $\eta$, via the Bogoliubov transformation introduced above. Suppose now that we take these evolved values as new annihilation and creation-like coefficients \emph{at time $\eta_{0}$}. This new choice then gives rise to a Fock representation of the Dirac field that is selected by a complex structure $\mathcal{J}_{\eta}$, which is clearly obtained from $\mathcal{J}_{\eta_{0}}$ by the dynamical evolution of the annihilation and creation-like variables \eqref{anni} and \eqref{creat}, from time $\eta_{0}$ to $\eta$. The requirement that we want to impose on the complex structures $\mathcal{J}_{\eta_{0}}$ and $\mathcal{J}_{\eta}$ is that they define unitarily equivalent Fock representations. This equivalence amounts to demand that the dynamics that relates the analyzed complex structures be implementable as a unitary operator on the Fock space defined, e.g., by $\mathcal{J}_{\eta_{0}}$, which is just the second condition in our criterion. Of course, in general, such dynamics is not exactly that of the coefficients $(e^{-3\alpha/2}x_{\vec{k}},e^{-3\alpha/2}\bar{y}_{\vec{k}})$, which is the evolution dictated by the Dirac equation \eqref{Deq}. This discrepancy is due to the time dependence of the functions $f_{l}^{\vec{k}}$ and $g_{l}^{\vec{k}}$. The explicit time dependence of these functions (and of the inverse power of the scale factor $e^{-3\alpha/2}$ appearing in the mode coefficients) absorbs part of the Dirac evolution. However, we will see that the part that is extracted is essentially fixed if we impose that the remaining dynamics be unitary. In other words, it is precisely because the functions $f_{l}^{\vec{k}}$ and $g_{l}^{\vec{k}}$ are explicitly time dependent that the dynamics that relates our family of annihilation and creation-like variables can be made unitarily implementable in the Fock quantum theory.

The Bogoliubov transformation that realizes the dynamics of the annihilation and creation-like variables \eqref{anni} and \eqref{creat} turns out to be implementable at any time $\eta$ as a unitary operator on the Fock space defined by $\mathcal{J}_{\eta_{0}}$ if and only if its antilinear part is Hilbert-Schmidt \cite{Shale,Derez}. This is equivalent to saying that the vacuum determined by the evolved annihilation and creation-like variables must contain a finite number of particles and antiparticles, according to the definition of particles and antiparticles of the initial vacuum. This happens if and only if the beta-coefficients of the Bogoliubov transformation are square summable at all times $\eta$:
\begin{align}\label{ucond}
\sum_{\vec{k}}|\beta_{\vec{k}}^{f}(\eta,\eta_0)|^{2}<\infty \qquad \text{and} \qquad \sum_{\vec{k}}|\beta_{\vec{k}}^{g}(\eta,\eta_0)|^{2}<\infty .
\end{align}
Thanks to relations \eqref{fgrel}, it is easy to see that $|\beta_{\vec{k}}^{f}(\eta,\eta_0)|=|\beta_{\vec{k}}^{g}(\eta,\eta_0)|$, so it suffices to analyze only one of these series. For that purpose, we will make use of \eqref{beta}, given for unspecified $h$, and we will take into account the asymptotic order of $h_{l}^{\vec{k}}$, for any $l=1,2$, in the limit of large $\omega_k$. The asymptotic order of the complementary coefficient $h_{\tilde{l}}^{\vec{k}}$ (where $\tilde{l}=2$ if $l=1$, and vice versa) is completely specified by relations \eqref{sympl}. 

Moreover, we will rule out any functional dependence of the functions $f_{l}^{\vec{k}}$ and $g_{l}^{\vec{k}}$ on the plane waves $e^{i\omega_{k}\Delta\eta}$ and  $e^{-i\omega_{k}\Delta\eta}$ that might absorb the dominant dynamical variation of the phases of $x_{\vec{k}}$ and $\bar{y}_{\vec{k}}$, in the regime of large values of $\omega_{k}$. The motivation to discard a behavior of this kind is to avoid considering trivial dynamics for the annihilation and creation-like variables in this asymptotic {\it ultraviolet} regime \cite{uf2}. Following a similar reasoning to that employed in \cite{uf2}, it is easy to check that, if $h_{l}^{\vec{k}}$ is negligible compared to $\omega_{k}^{-1}$ for all tuples $\vec{k}$ in any infinite sublattice of $\mathbb{Z}^{3}$, then conditions \eqref{ucond} can never be met. Thus, at least for $\eta$ in a sufficiently small interval beyond $\eta_{0}$, we are left with the (mutually non-exclusive) possibilities that $h_{l}^{\vec{k}}$ is either of the order of $\omega_{k}^{-1}$ or it is bigger, for all tuples $\vec{k}\in\mathbb{Z}^{3}$, up to a possible and irrelevant finite subset of tuples. In these circumstances, one can carefully derive the necessary and sufficient conditions on $h_{l}^{\vec{k}}$ for \eqref{ucond} to hold, without introducing a trivialization of the dynamics in the sense explained above. Denoting $\{l,\tilde{l}\}=\{1,2\}$ as a set, these conditions are the following.

\begin{itemize}
\item[i)] There may exist an infinite sublattice $\mathbb{Z}_{l,\uparrow}^{3}\subset \mathbb{Z}^{3}$ of tuples $\vec{k}$ such that the functions $h_{l}^{\vec{k}}$, which must be asymptotically of order $\omega_{k}^{-1}$ or higher, form a sequence that is square summable at all times $\eta$. This condition, in particular, implies that $h_{l}^{\vec{k}}$ must tend to zero when $\omega_{k}$ tends to infinity for tuples in $\mathbb{Z}_{l,\uparrow}^{3}$. It then follows from \eqref{sympl} that, in the sublattice $\mathbb{Z}_{l,\uparrow}^{3}$, the other function $h^{\vec{k}}_{\tilde{l}}$ must be asymptotically of order unity plus terms $\mathcal{O}(|h^{\vec{k}}_{l}|^{2})$.

\item[ii)] For all tuples $\vec{k}\in\mathbb{Z}_{l}^{3}$, where $\mathbb{Z}_{1}^{3}\cup\mathbb{Z}_{2}^{3}$ is the complementary sublattice of $\mathbb{Z}_{1,\uparrow}^{3}\cup\mathbb{Z}_{2,\uparrow}^{3}$ in $\mathbb{Z}^{3}$ (up to a finite number of elements), the functions $h_{l}^{\vec{k}}$ satisfy
\begin{align}\label{unith}
h^{\vec{k}}_{l}=(-1)^{l+1}\frac{me^{\alpha}}{2\omega_{k}}e^{iH^{\vec{k}}_{\tilde{l}}}+\vartheta_{h,l}^{\vec{k}}
\end{align}
for all times $\eta$. It may happen that one of the sublattices $\mathbb{Z}_{l}^{3}$ is void. Here, $H^{\vec{k}}_{\tilde{l}}$ is the phase of $h^{\vec{k}}_{\tilde{l}}$, possibly time dependent, and $\vartheta_{h,l}^{\vec{k}}$ denotes subdominant terms (negligible compared to $\omega_{k}^{-1}$) such that
\begin{align}
\sum_{\vec{k}\in \mathbb{Z}_{l}^{3}}\left\vert\vartheta_{h,l}^{\vec{k}}\right\vert^{2}<\infty, \qquad \forall\eta.
\end{align}
\end{itemize}

As above, we have followed here arguments similar to those of \cite{uf2}. In particular, we have used that the sequence $\{\omega_{k}^{-2}\}_{\vec{k}\in\mathbb{Z}^{3}}$ is square summable, despite the growing asymptotic behavior of the degeneracy of the Dirac eigenvalues on $T^3$ (see e.g. \cite{torus1} and notice that here $\omega_{k}$ grows asymptotically as the norm of $\vec{k}$, regardless of the choice of spin structure).

In conclusion, any Fock representation of the CARs of the Dirac field defined by \eqref{anni} and \eqref{creat} admits a non-trivial unitarily implementable dynamics if and only if the time-dependent functions $h^{\vec{k}}_{l}$ satisfy conditions i) and ii) above. Let us emphasize that at least one of the sublattices $\mathbb{Z}_{l}^{3}$, in which $h^{\vec{k}}_{l}$ has to be of the form \eqref{unith}, must have an infinite number of elements. In fact, if this were not the case, one can see that condition i) could not be met ($\mathbb{Z}_{l,\uparrow}^{3}$ would have too many elements, preventing the square summability of $h_{l}^{\vec{k}}$ on this sublattice, because $h_{l}^{\vec{k}}$ is assumed to be of order $\omega_k^{-1}$ or higher there, and the degeneracy $g_k$ grows faster than $\omega_k$). As a result, the requirement of unitarily implementable dynamics fixes completely the explicit time dependence of the dominant parts of the annihilation and creation-like variables \eqref{anni} and \eqref{creat} for an infinite number of modes in the asymptotic regime of large Dirac eigenvalues. That is to say, the time-dependent functions that need to be extracted from the Dirac dynamics, in order that the remaining evolution can be implemented as a unitary operator, are completely specified at dominant order (in norm). Specifically, they are given by the product of the factor $me^{\alpha}/2\omega_{k}$ in \eqref{unith} and the factor $e^{3\alpha/2}$, introduced in the definition of the time-dependent coefficients $(x_{\vec{k}},\bar{y}_{\vec{k}})$ that span $\Psi$.

\section{Uniqueness of the Fock representation}
\label{sec:uniqueness}

Among all the invariant complex structures for the Dirac field in the flat FLRW cosmology with compact sections, we now know which families of them are related by a non-trivial fermion dynamics that is implementable as a unitary operator on Fock space. It is of course desirable to elucidate if these families of complex structures, and therefore of Fock representations, are all unitarily equivalent or not. If the answer to this question is in the affirmative, then our criterion of invariance under the physical symmetries of the Dirac-FLRW system and of a unitarily implementable dynamics indeed selects a unique vacuum for the Dirac field, up to unitary equivalence. Moreover, in that case the quantum dynamics of the annihilation and creation operators is essentially fixed, inasmuch as any allowed redefinition of the quantum evolution does not affect its dominant part (in the ultraviolet region) and besides is a unitarily implementable transformation.

In order to attain this uniqueness result, let us consider two families of invariant complex structures: a fixed one, denoted by $\mathcal{J}_{R}$, that serves as reference, and any arbitrary one, $\tilde{\mathcal{J}}$, within the class picked out by our previous requirements. We use here a simplified notation which does not display the $\eta$-dependence of the elements of these families of complex structures (nonetheless, this dependence is clear in all the discussion). Our reference vacuum is characterized by annihilation and creation-like variables $a_{\vec{k}}^{(x,y)}$ and $b_{\vec{k}}^{\dagger (x,y)}$ of the form \eqref{anni} and \eqref{creat}, with
\begin{align}\label{ref}
f_{1}^{\vec{k}}=\frac{me^{\alpha}}{2\omega_{k}}, \qquad f_{2}^{\vec{k}}=\sqrt{1-\frac{m^{2}e^{2\alpha}}{4\omega^{2}_{k}}}, \qquad g_{1}^{\vec{k}}=f_{2}^{\vec{k}},\qquad g_{2}^{\vec{k}}=-f_{1}^{\vec{k}}
\end{align}
for all $\vec{k}\in\mathbb{Z}^{3}$. On the other hand, the Fock quantization selected by $\tilde{\mathcal{J}}$ will be determined by similar annihilation and creation-like variables $\tilde{a}_{\vec{k}}^{(x,y)}$ and $\tilde{b}_{\vec{k}}^{\dagger (x,y)}$, but with (smooth) time-dependent functions $\tilde{f}_{l}^{\vec{k}}$ and $\tilde{g}_{l}^{\vec{k}}$ (with $l=1,2$) that are only subject to the restriction that either  $\tilde{f}_{l}^{\vec{k}}$ or alternatively $\tilde{g}_{l}^{\vec{k}}$ must satisfy the non-trivial unitary dynamics conditions i) and ii), explained in the previous section. For concreteness, we will focus on the case in which $\tilde{f}_{l}^{\vec{k}}$ is the function that fulfills those conditions [the other situation can be dealt with in a completely analogous way, owing to relations \eqref{fgrel}]. 

The relation between the two considered sets of annihilation and creation-like variables at any instant of time $\eta$ is a Bogoliubov transformation. Let us call $\lambda_{\vec{k}}^{f}(\eta)$ and $\lambda_{\vec{k}}^{g}(\eta)$ the coefficients of this transformation that mix the creation and annihilation parts of the field, relating $\tilde{a}_{\vec{k}}^{(x,y)}(\eta)$ with $b_{\vec{k}}^{\dagger (x,y)}(\eta)$ and $\tilde{b}_{\vec{k}}^{\dagger (x,y)}(\eta)$ with $a_{\vec{k}}^{(x,y)}(\eta)$, respectively. It was shown in \cite{uf1} that the norm of these coefficients is given by 
\begin{align}\label{lambda}
\left\vert \lambda^{h}_{\vec{k}}\right\vert=\left\vert\tilde{h}_{1}^{\vec{k}}h_{2}^{\vec{k}}-\tilde{h}_{2}^{\vec{k}}h_{1}^{\vec{k}}\right\vert,
\end{align}
where we recall that $h=f,g$. Again, the families of Fock representations defined by $\mathcal{J}_{R}$ and $\tilde{\mathcal{J}}$ will be unitarily equivalent (and hence also the two corresponding dynamics) if and only if the sequences formed by $\lambda^{h}_{\vec{k}}$, with $\vec{k}\in\mathbb{Z}^{3}$, are square summable for all $\eta$. Let us note that relations \eqref{fgrel} guarantee again that $\left\vert\lambda^{f}_{\vec{k}}\right\vert=\left\vert\lambda^{g}_{\vec{k}}\right\vert$, so it suffices to analyze only one of these sequences of coefficients. It is easy to check that
\begin{align}\label{uequiv}
\left\vert\lambda^{f}_{\vec{k}}\right\vert=\left\vert \tilde{f}_{1}^{\vec{k}}\right\vert+o\left(\tilde{f}_{1}^{\vec{k}}\right) \quad \forall\vec{k}\in\mathbb{Z}_{1,\uparrow}^{3}, \qquad \left\vert\lambda^{f}_{\vec{k}}\right\vert=\left\vert\vartheta_{\tilde{f},1}^{\vec{k}}\right\vert+\mathcal{O}(\omega_{k}^{-2}) \quad \forall\vec{k}\in\mathbb{Z}_{1}^{3},
\end{align}
whereas $\left\vert\lambda^{f}_{\vec{k}}\right\vert=\mathcal{O}(1)$ for all $\vec{k}\in\mathbb{Z}_{2,\uparrow}^{3}\cup\mathbb{Z}_{2}^{3}$. Here, the symbol $o(.)$ stands for terms that are negligible with respect to its argument. Therefore, the lambda coefficients that relate $\mathcal{J}_{R}$ with $\tilde{\mathcal{J}}$ at any instant of the conformal time are clearly square summable in the infinite subset $\vec{k}\in\mathbb{Z}_{1,\uparrow}^{3}\cup\mathbb{Z}_{1}^{3}$, although the square summability is lost in the complementary subsequence for $\vec{k}\in\mathbb{Z}_{2,\uparrow}^{3}\cup\mathbb{Z}_{2}^{3}$ (unless this sublattice is chosen void). However, using relations \eqref{fgrel}, one can argue that if the roles of $\tilde{f}_{l}^{\vec{k}}$ and $\tilde{g}_{l}^{\vec{k}}$ are interchanged for all $\vec{k}\in\mathbb{Z}_{2,\uparrow}^{3}\cup\mathbb{Z}_{2}^{3}$, then the resulting lambda coefficients are no longer of order unity, but they turn out to be square summable indeed. In this way, the complex structure resulting from $\tilde{\mathcal{J}}$ by means of this interchange defines a quantum theory which is unitarily equivalent to the one defined by $\mathcal{J}_{R}$.

Given relations \eqref{anni} and \eqref{creat}, the interchange $\tilde{f}_{l}^{\vec{k}}\leftrightarrow\tilde{g}_{l}^{\vec{k}}$ that we have employed above obviously corresponds to an interchange between the notion of  particles and the notion of antiparticles for the (possibly) infinite number of degrees of freedom corresponding to $\mathbb{Z}_{2,\uparrow}^{3}\cup\mathbb{Z}_{2}^{3}$. In this sense, one can regard all the possible splittings of the lattice $\mathbb{Z}^{3}$, into the infinite sublattices $\mathbb{Z}_{1,\uparrow}^{3}\cup\mathbb{Z}_{1}^{3}$ and $\mathbb{Z}_{2,\uparrow}^{3}\cup\mathbb{Z}_{2}^{3}$ (with one of them being possibly void), as all the distinct choices of conventions for the definition of particles and antiparticles within the class of invariant Fock representations with a non-trivial unitarily implementable dynamics. Once one of such conventions is fixed, our discussion proves that all the vacua allowed by our criterion are unitarily equivalent. The question remains of whether there exists a physical condition that selects a natural convention. In order to provide an answer, it is enlightening to notice that $f_{1}^{\vec{k}}$ and $f_{2}^{\vec{k}}$ quantify, respectively, the left and right-handedness (under chirality) of the particles associated with $a_{\vec{k}}^{(m,s)}$. And precisely the opposite happens with the time-dependent functions that characterize $a_{\vec{k}}^{(t,r)}$. Besides, we recall that the set of coefficients $\{m_{\vec{k}},\bar{s}_{\vec{k}}\}$ capture the positive helicity of the Dirac field, whereas $\{t_{\vec{k}},\bar{r}_{\vec{k}}\}$ capture the negative helicity. One could then impose that a physically reasonable quantization of the Dirac field must be such that, at a given time $\eta$, all the excitations that correspond to particles of a given (and conserved) helicity display the same relative chirality. The corresponding antiparticle excitations would then display essentially the opposite chirality, owing to relations \eqref{fgrel}. Such restriction on the allowed vacua would be translated into the condition that the lattice $\mathbb{Z}^{3}$ must equal either $\mathbb{Z}_{1,\uparrow}^{3}\cup\mathbb{Z}_{1}^{3}$, or $\mathbb{Z}_{2,\uparrow}^{3}\cup\mathbb{Z}_{2}^{3}$. This two possible choices of convention for the notions of particles and antiparticles may be further reduced to one if one requires a smooth transition from the convention defined for the massive field to the limit in which the mass tends to zero, limit where the chirality becomes a conserved quantity. Indeed, if one demands this smoothness and that the massless particle excitations of positive helicity are, e.g., dominantly right-handed, while those of negative helicity are dominantly left-handed, one clearly restricts all considerations to the convention that assigns $\mathbb{Z}^{3}= \mathbb{Z}_{1,\uparrow}^{3}\cup\mathbb{Z}_{1}^{3}$, both for $a_{\vec{k}}^{(m,s)}$ and $a_{\vec{k}}^{(t,r)}$. Setting this convention, the Fock vacuum characterized by \eqref{ref} is the simplest choice allowed by our criterion of symmetry invariance and of a non-trivial unitary dynamics.

In summary, if we take the same convention for particles and antiparticles adopted with the choice of family of complex structures $\mathcal{J}_{R}$, then we have proven that, up to unitary equivalence, the Fock representation of the CARs that this family defines is the only invariant representation compatible with the existence of a non-trivial quantum dynamics that is implementable by a unitary operator on Fock space.

\section{Discussion}
\label{sec:conclu}

In this work, we have given a criterion for the choice of a Fock representation of the CARs of a Dirac field propagating in an FLRW cosmology with flat and compact spatial sections (isomorphic to tori), and we have proven the uniqueness of this choice up to unitary equivalence. Moreover, the criterion provides an essentially unique splitting of the field time dependence into an explicit dependence on the background and a genuine quantum dynamics. The result is attained by imposing two physically admissible and general enough conditions, and once a convention for the notions of particles and antiparticles has been adopted. The fundamental condition imposed on the Fock quantization is that it admits a non-trivial quantum evolution that is implementable as a unitary operator on Fock space. Complementing this unitarity condition, we have required that the quantum theory must be naturally invariant both under the group of Killing symmetries of the spatial sections of the cosmology and under the group of spin rotations generated by the conserved helicity of the Dirac field. In addition, we have adopted a natural convention for the concepts of particles and antiparticles, distinguished by imposing consistency in the chirality of these two types of excitations at arbitrary time, together with the demand of a smooth transition to the conventional behavior found in the massless limit.

In order to handle the system more easily, we have expressed each of the two-component spinors of opposite chirality that describe the Dirac field in terms of a basis (and its complex conjugate) of eigenspinors of the Dirac operator on $T^{3}$, with time-dependent coefficients. A key point in our demonstration of uniqueness has been the knowledge about the asymptotic behavior of these coefficients for solutions of the Dirac equation, in the limit of large Dirac eigenvalues and up to a sufficiently high order \cite{uf1}. With such information at hand, we have been able to characterize the general form of the annihilation and creation-like variables associated with invariant vacua and such that the dynamics of those variables is unitarily implementable. Employing this characterization and taking the above suitable convention for the notions of particles and antiparticles, we have derived the uniqueness result almost straightforwardly. More concretely, we have shown that the field dynamics that follows from the Dirac equation can be implemented as a non-trivial unitary operator on Fock space only after extracting certain time variation from its dominant part (with respect to the limit of large Dirac eigenvalues). This time variation has been captured in a series of functions, that are  explicitly dependent on time or, equivalently, on the scale factor of the cosmology under study, and that, at the end of the day, determine the definition of the annihilation and creation-like variables of the Fock quantum theory. Actually, such time dependence is different when the field is massive and when is massless. This distinction between the massive and massless cases arises owing to the dynamical coupling between the two chiralities of the Dirac field, which occurs when the mass is non-zero, and that makes the chirality a non-conserved property. Indeed, the coupling implies that the massive contribution to the the solutions of the Dirac equation cannot be ignored in the limit of large eigenvalues of the Dirac operator. This contribution involves the scale factor of the FLRW cosmology, a time-varying function. This explicit time dependence needs then to be extracted in order for the remaining dynamics to be unitarily implementable.

Our results thus characterize completely which particle and antiparticle excitations of the Dirac field can undergo a non-trivial evolution that respects the quantum coherence over time. Besides, the background dependence of the field that completes this unitary quantum transformation into the full Dirac dynamics is also totally specified, as commented above. This determination of the excitations that pose no loss of information in the quantum evolution may be useful in contexts involving particle detectors, or even when surpassing the framework of quantum field theory in curved spacetimes and considering the cosmological system as a fully quantum entity.

\section*{Acknowledgments}
B. Elizaga Navascu\'es is grateful to C. Barcel\'o for helpful conversations. She also recognizes the hospitality provided by the High Energy Physics Department of the Radboud University Nijmegen during the months of April and May, 2016. This work was partially supported by the research grants MINECO Project No. FIS2014-54800-C2-2-P from Spain, DGAPA-UNAM IN113115 and CONACyT 237351 from Mexico, and COST Action MP1405 QSPACE, supported by COST (European Cooperation in Science and Technology). In addition, M. M-B acknowledges financial support from the Netherlands Organisation for Scientific Research (NWO).

\end{document}